# Application analysis of ai technology combined with spiral CT scanning in early lung cancer screening

Shulin Li [1,*], Liqiang Yu[2], Bo Liu [3], Qunwei Lin[4], Jiaxin Huang[5]

[1] Information Studies,Trine University,Phoenix, USA
[2] Computational Social Sciences,The University of Chicago,Irvine, USA
[3] Software Engineering,Zhejiang University,Hangzhou China
[4] Information Studies,Trine University,Phoenix,USA
[5] Information Studies,Trine University,Phoenix,USA

*  **Corresponding author**:  Shulin Li(Email: liam.cool666@gmail.com)

**Abstract:** At present, the incidence and fatality rate of lung cancer in China rank first among all malignant tumors. Despite the continuous development and improvement of China's medical level, the overall 5-year survival rate of lung cancer patients is still lower than 20% and is staged. A number of studies have confirmed that early diagnosis and treatment of early stage lung cancer is of great significance to improve the prognosis of patients. In recent years, artificial intelligence technology has gradually begun to be applied in oncology. ai is used in cancer screening, clinical diagnosis, radiation therapy (image acquisition, at-risk organ segmentation, image calibration and delivery) and other aspects of rapid development. However, whether medical ai can be socialized depends on the public's attitude and acceptance to a certain extent. However, at present, there are few studies on the diagnosis of early lung cancer by AI technology combined with SCT scanning. In view of this, this study applied the combined method in early lung cancer screening, aiming to find a safe and efficient screening mode and provide a reference for clinical diagnosis and treatment.

## 1. INTRODUCTION

It has been reported that Spiral computed tomography (SCT) is more helpful to improve the detection rate of early lung cancer than chest X-ray. Compared with conventional CT, SCT can clearly display the lung structure of patients without superimposed radiation amount, but due to the small size of nodules, the proximity of the central location to blood vessels, and the lack of clarity at the edges, there are still some missed diagnosis

Risk. At the same time, a large number of film reading can also cause visual and mental fatigue of imaging doctors, leading to clinical missed diagnosis and misdiagnosis. Artificial Intelligence (AI) technology is a branch of computer science, and its application in imaging examination can effectively assist doctors in the diagnosis of film reading and reduce clinical errors. However, at present, there are few researches on the diagnosis of early lung cancer by AI technology combined with SCT scanning. In view of this, this study applied the combined method in early lung cancer screening, aiming to find a safe and efficient screening mode and provide a reference for clinical diagnosis and treatment.

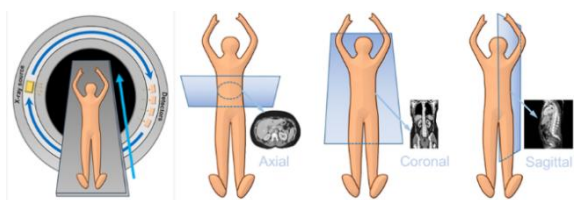

**Figure 1.** Spiral CT operation process

CT (computed tomography) uses X-rays to obtain images. A heated cathode releases high-energy electrons, which in turn release their energy as X-ray radiation. X-rays pass through tissues and hit a detector on the other side. The more dense a tissue, the more X-rays it absorbs.

Bone: X-rays absorbed = few X-rays reaching detector: White

Air: X-rays not absorbed = lots of X-rays reaching detector: Black Compared to plain film, **CT is able to distinguish more subtle density differences and there is no overlap of structures.

Current CT machines use 'Spiral CT'. This consists of a single radiation source with multiple detectors which rotates around the patient, obtaining a block of data as the patient is moved through.

## 2. RELATED WORK

Pulmonary nodule disease is an allergic disease of unknown cause with pathological manifestations of non-cheesy meat. Granuloma can invade organs and tissues of the whole body, and the organs commonly involved are lymph nodes, lungs, liver, spleen and skin, etc[1]. The high incidence of pulmonary nodule lesions occurs in people aged 10-40 years, and the peak incidence occurs in people aged 20-30 years, and there is a gender difference, and it is more common in women. The identification of pulmonary nodule characteristics is of auxiliary significance for early screening of lung cancer and has positive significance for improving the survival time of lung cancer patients. According to the nature of pulmonary nodular lesions[2].

It can be divided into benign and malignant nodules, and the treatment methods of different nodule properties vary greatly, so it is of great significance to clarify the nature of nodules[3-4]. Wang Wei et al. pointed out that due to the small diameter of pulmonary nodules, routine examination is not of high diagnostic value, which is prone to misdiagnosis and missed diagnosis, causing delay in early treatment of the disease and affecting the quality of life of patients. Chest

X-ray and sputum cytology have been commonly used in clinical diagnosis of pulmonary nodules, but the diagnostic effect is not good through large sample screening. With the development of multi-slice spiral CT post-processing technology, CT examination has been gradually used by clinicians to assist the diagnosis of pulmonary nodules. It has been reported abroad that CT examination can quickly demarcate abnormal areas, accurately identify benign and malignant nodules, measure the volume of nodules and observe the shape of nodules[5-8]. Due to the limitation of thickness and scanning speed, CT plain scan is not effective in the early diagnosis of lung cancer[8-9]. 16-row low-dose CT scan is of high value in the detection of small pulmonary nodules, which can quickly perform a wide range of scanning and multi-plane thin-layer reconstruction, and clearly show the small pulmonary nodules[10]. However, there are relatively few such reports in China, so in this paper, spiral CT and 3D reconstruction are used in the diagnosis of pulmonary nodules and their application value is observed.

## 3. Methodology

Windowing, also known as grey-level mapping, contrast stretching, histogram modification or contrastenhancement is the process in which the CT image greyscale component of an image is manipulated via the CT numbers; doing this will change the appearance of the picture to highlight particular structures. The brightness of the image is adjusted via the window level. The contrast is adjusted via the window width.

Tissue density is measured in Hounsfield units (HU).
This is defined as c **Air = −1000 HU; Water = 0 HU.**
Density of tissues in CT-Scans:

$$Air < Fat < Fluid < Soft tissue < Bone < Metal \qquad (1)$$

The easier way to remember this is (Fat floats on water, so is less dense than fluid; Soft tissue is mostly intracellular fluid with some connective tissue).

### 3.1. Model Architectures

The window width (WW) as the name suggests is the measure of the range of CT numbers that an image contains. A wider window width (2000 HU), therefore, will display a wider range of CT numbers. Consequently, the transition of dark to light structures will occur over a larger transition area to that of a narrow window width (<1000 HU). Accordingly, it is important to note, that a significantly wide window displaying all the CT numbers will result in different attenuations between soft tissues to become obscured.

**Wide window:** When you are looking at an area with predominantly different tissue density, a wide window is used. A good example is lungs or cortical tissue, where air and vessels will sit side by side.

**Narrow window:** When you are looking at tissues with almost similar density, you should use narrow window. As a result subtle changes in tissued density (small window) is magnified over the whole grayscale range.

Window level/center

The window level (WL), often also referred to as window center, is the midpoint of the range of the CT numbers displayed. When the window level is decreased the CT image will be brighter and vice versa.

When presented with a Window width (WW) and Widnow Level (WL) one can calculate the upper and lower grey levels i.e. values over x will be white and values below y will be black.

the upper grey level (x) is calculated via WL + (WW ÷ 2)
the lower grey level (y) is calculated via WL - (WW ÷ 2)

For example, a brain is W:80 L:40. Therefore, all values above +80 will be set to maximum grayscale level (white) and all values below 0 will be set to lowest grayscale level in the display (black). And values between +0 to +80 will be spread between the whole grayscale range.

Examples of commonly used windows are soft tissue, lung, and bone are given below:

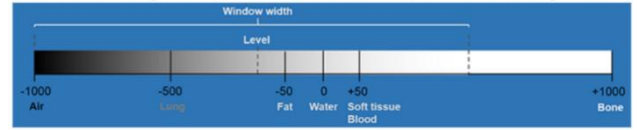

**Figure 2.** Level: −200 HU; Width: 2000 HU (Range: −1200 to +800),

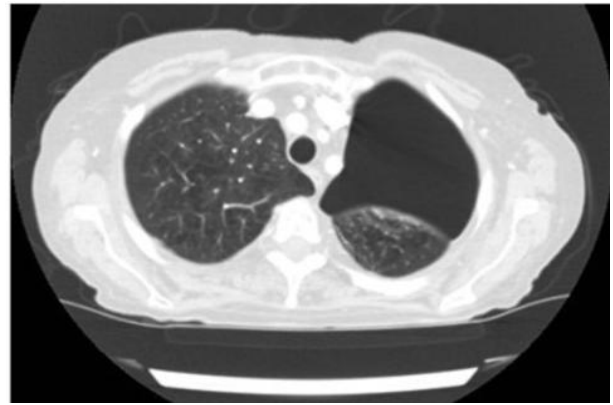

**Figure 2**. Lung window on a chest CT

Multi-slice spiral CT (Siemens SOMATOMScope) was used to scan, and the parameters were: tube current 50mAs, tube voltage 120kV, layer thickness and layer spacing were 5mm, pitch was 1.0, and rotation period was 0.5s. The patient was placed in the supine position and scanned from the thoracic entrance to the costophrenic Angle. CT scan was performed, and then iodohexyl (90mL) was injected through the cubital vein with a high pressure syringe at a rate of 4.0mL/s. Transfer the image to the workstation for Volume surface projection display (SSD), Multiplanar reconstruction, MPR, Volume of interest method (VOI) three-dimensional reconstruction inspection.

### 3.2. DataSet Introduction

The results of pathological examination are the gold standard. The diagnostic criteria of malignant nodules were as follows: burr or lobulated edge contour, pleural pull, vacuol-like internal tumor, and bronchial gas phase. In the absence of these signs, the nodules are benign.

$$\frac{\partial u}{\partial t} = \nabla^2 u + f(u, v),$$
$$\frac{\partial v}{\partial t} = \delta \nabla^2 v + g(u, v),$$
(1)

$$f(u, v) = \frac{1}{\epsilon} u(1 - u)\left(u - \frac{v + b}{a}\right).$$
(2)

There is freedom in the choice of g(u,v), and our methods will not depend on this choice. However, the results we report will be for the g proposed by Bar and Eiswirth and used by Sandstede and Scheel , namely,

$$g(u,v) = \begin{cases} -v, & 0 \leq u < 1/3, \\ 1 - 6.75u(u-1)^2 - v, & 1/3 \leq u \leq 1, \\ 1 - v, & 1 < u. \end{cases}$$
(3)

The equations are posed on a disk of radius R and with homogencous Neumann boundary conditions at r = R:

$$\frac{\partial u}{\partial r}(R, \theta) = \frac{\partial v}{\partial r}(R, \theta) = 0,$$
(4)

where r,0 are standard polar coordinates. For chemically reacting systems these are the most natural boundary conditions, as they correspond to zero chemical fux through the boundary of the domain. Other boundary conditions could give different spiral solutions and linear stability spectra on finite domains, but we do not consider any other boundary conditions here.

CT showed hilar and mediastinal lymph node enlargement, and the rule of lymph node metastasis followed the rule of metastasis from the lung to the mediastinum through the hilar. Therefore, ipsilateral hilar and mediastinal lymph node enlargement was common, and a few only showed hilar or mediastinal lymph node enlargement

Mediastinal lymph node enlargement, some may appear bilateral hilar and mediastinal lymph node metastasis.

## 4. Conclusion

Pulmonary nodular disease is a common lung disease. Recently, the incidence of pulmonary nodular disease is increasing. Pulmonary nodular lesions can be classified into benign and malignant lesions according to pathological properties. However, due to the small diameter of the nodules, most of which are less than 3cm, routine examination is of little value in distinguishing the nature of pulmonary nodules, and the probability of misdiagnosis and missed diagnosis is high. There is a great difference in the treatment of benign nodules and malignant nodules[11]. Therefore, it is of great significance to find a method to accurately identify the nature of pulmonary nodules to formulate treatment plan and improve the outcome of the disease. CT is a common means of clinical diagnosis of pulmonary nodular lesions, which is convenient to operate, non-invasive to the body, fast scanning speed, and strong repeatability, so it is favored by clinicians and patients. Spiral CT is a commonly used diagnostic method for clinical diagnosis of pulmonary nodular lesions. The image and time resolution of this scan are both high, which can provide physicians with morphological information of nodules, help physicians observe the pathophysiology of nodules, and provide great help for clinicians in diagnosis[12]. However, spiral CT plain scan can not clearly show the direct relationship between bronchial arteries and some lesions, which is not conducive to pulmonary nodular diseases

Become diagnosed. In the process of diagnosing small pulmonary nodules, low-dose CT scan can quickly scan the overlapping lesions of mediastinum and diaphragm, and perform high-resolution multi-plane reconstruction to obtain clear images of small pulmonary nodules. As a new imaging method, 3D reconstruction technology is gradually used in clinical diagnosis of diseases. This technology can use software to post-process images and obtain clear diseases[13].

The stove image. SSD, MPR and VOI are the common methods of 3D reconstruction technology. The above methods can obtain the oblique plane of the section fault at any Angle by using the computer to reassemble any section. The image processed by 3D reconstruction technology has no artifacts, and the three-dimensional shape of the lesion and the spatial relationship between the lesion and neighboring tissues can be clearly observed, which is conducive to doctors to make accurate diagnosis.

In summary, 3D reconstruction technology after multi-slice spiral CT image has certain application value in the diagnosis of benign and malignant pulmonary nodules of fox elevation, but there is still a high rate of missed diagnosis. Parallel test using CT plain scan and 3D reconstruction technology can greatly improve the sensitivity and consistency of clinical diagnosis and has good clinical application value. However, the sample size of the included cases in this study is small and the selection is limited, so it is necessary to increase the sample size and conduct more in-depth research in the future.